# Carbon Capture and Separation from $CO_2/N_2/H_2O$ Gaseous Mixtures in Bilayer Graphtriyne: A Molecular Dynamics Study

Noelia Faginas-Lago [a], *[0000-0002-4056-3364], Yusuf Bramastya Apriliyanto [b][0000-0003-0683-8456], Andrea Lombardi [a] [0000-0002-7875-2697]

[a] Dipartimento di Chimica, Biologia e Biotecnologie, Università degli Studi di Perugia, Via Elce di Sotto 8, 06123, Perugia, Italy

[b] Department of Chemistry, IPB University, Jl Tanjung Kampus IPB Dramaga, 16680, Bogor, Indonesia

**Abstract**

Molecular dynamics simulations have been performed for $CO_2$ capture and separation from $CO_2/N_2/H_2O$ gaseous mixtures in bilayer graphtriyne. The gas uptake capacity, permeability as well as selectivity of the layers were simulated based on an improved formulation of force fields tested on accurate *ab initio* calculations on specific systems for mixture separation in post-combustion process. The effect of pressure and temperature on the separation performances of graphtriyne layers was investigated. Compared with the single layer graphtriyne, bilayer graphtriyne can adsorb more molecules with relatively good selectivity, due to the action of the interlayer region as an adsorption site. The interlayer adsorption selectivity of $CO_2/N_2$ and $CO_2/H_2O$ at a temperature of 333 K and a pressure of 4 atm have been found to be equal to about 20.23 and 1.85, respectively. We also observed that the bilayer graphtriyne membrane has high $CO_2$ and $H_2O$ permeances compared to $N_2$ with permeance selectivity ranging from 4.8 to 6.5. Moreover, we found that permeation and adsorption depend on the applied temperature; at high temperatures permeation and adsorption tend to decrease for all molecules.

## 1. Introduction

Gas adsorption by porous materials as a phenomenon to ground technologies aimed at reducing greenhouse gas emissions has actively been investigated in the past few years[1-3]. All of these efforts are eventually directed to mitigate possible anthropic contributions to climate change and global warming. Among the greenhouse gases, $CO_2$ is the most abundant and is regularly released into the atmosphere [4]. Therefore, post-combustion $CO_2$ capture and separation are important to control the $CO_2$ emission mainly produced from fossil fuels combustion. Major advantages of porous materials over the traditional aqueous chemical absorbent are in terms of its simplicity, recovery and low implementation costs [5]. Carbon-based materials bearing intrinsic pores currently emerge as potential solid adsorbent candidates for $CO_2$ capture [6-8]. Unlike other porous materials (e.g. MOFs and polymers), carbon-based materials are hydrophobic, chemically inert and thermally stable. Thanks to their exceptional properties, carbon-based materials are not susceptible to heat and water vapour, which is a characteristic of post-combustion flue gas. Moreover, compared with MOFs and zeolites, carbon-based materials are constructed from lightweight elements linked by strong covalent bonds producing low density and robust structures [9-13]. Therefore, carbon-based materials are economically suitable and viable for post combustion $CO_2$ capture and separation.

Instead of only capture $CO_2$ molecules by surface adsorption, two-dimensional carbon-based membranes provide a unique properties harnessed from a combination of surface adsorption and their intrinsic pores acting as molecular sieving to separate $CO_2$ from other gaseous mixtures [14,15]. Graphene, a class of carbon-based materials, have been attracting much attention in recent years. It is a single atom thick planar membrane with remarkable properties [16-18]. Surface functionalization by introducing holes or attaching different organic groups is commonly performed in order to modify graphene to meet the requirements for a specific application. For gas adsorption and separation, nano-porous graphene is reported as a promising porous membrane material based on carbon [19,20]. However, a major drawback of nano-porous graphene is the tendency to form aggregates, that limits its gas uptake capacity. Moreover, practically, it is difficult to control the pore size and the homogeneity of their distribution. Analogous with the graphene, γ-graphynes are single atomic layers belonging to the class of carbon allotropes. The carbon atoms in γ-graphynes are arranged as a function of the C-C triple bonds bridging two adjacent hexagons. Possessing similar properties with graphene, γ-graphynes have uniformly distributed and adjustable pores [21-23]. In addition, with lower dispersion forces and therefore a reduced tendency to form aggregates, γ-graphynes are suited for gas capture and separation [24,25].

Since synthesis and characterization techniques of graphynes are still actively being developed [26-29], computer modelling and simulations play an important role in the material development and evaluation of their performance for gas capture and separation [30,31]. Recent studies reported that graphtriyne (a form of graphynes, in which each benzene ring is connected to each of six others through a chain composed by three acetylenic bonds), showed a strong physisorption of $CO_2$ over $N_2$ and $H_2O$ molecules [32,33]. It is also reported that the adsorption energy of $CO_2$ can be enhanced by introducing a new layer of graphtriyne over an existing graphtriyne membrane and so on. The multi-layered graphtriyne systems introduce interlayer spaces that can confine $CO_2$ molecules trough strong attractive forces. These interlayer spaces can be exploited as new adsorption sites thus the uptake capacity of $CO_2$ can be increased [34]. In this work, a study of the performance of bilayer graphtriyne membrane for $CO_2$ capture and separation is evaluated using extended molecular dynamics (MD) simulations. A wide range of relevant conditions in post combustion involving a gaseous mixture of $CO_2/N_2/H_2O$ has been applied in the molecular simulations.

## 2. Methods

For modelling the intermolecular interactions, the total potential energy of the system is split into electrostatic and non-electrostatic contributions. The total potential energy represents the sum of the intermolecular potential energy between any interacting pair in the system involving $CO_2$, $N_2$, $H_2O$ and graphtriyne layers. The electrostatic contribution is calculated using a standard Coulombic summation, by assigning point charges to the interacting molecules. Meanwhile, the non-electrostatic term is expressed using the Improved Lennard-Jones (ILJ) potential [35], as follows:

$$V_{tot}(r) = \sum_{i,j}^{n} \frac{q_i q_j}{r_{ij}} + \sum_{i}^{n} V_{ILJ}(r_i)$$

where

$$V_{ILJ}(r_i) = \varepsilon \left[ \frac{m}{n(r)-m} \left(\frac{r_0}{r}\right)^{n(r)} - \frac{n(r)}{n(r)-m} \left(\frac{r_0}{r}\right)^{m} \right]$$

with

$$n(r) = \beta + 4.0 \left(\frac{r}{r_0}\right)^2$$

As can be seen from the above equations, the ILJ potential requires four parameters ($r_0$, $\varepsilon$, $m$ and $\beta$) to be specified. The $r_0$ and $\varepsilon$ are pair specific parameters representing the equilibrium distance and the depth of the potential well, respectively. The $m$ parameter takes the value of 6 for describing interactions between neutral molecules, while the dimensionless $\beta$ is a parameter that can be adjusted following the hardness of the interaction. The $\beta$ parameter also corrects the dependence of the interaction on the internuclear distance, improving the potential function at the asymptotic region. The ILJ function so formulated versatile improves the energy profile at equilibrium distance and at short and long range compared to the traditional Lennard-Jones (LJ) function. For a full account of the advantages of the ILJ function see Refs [36-42] and references therein. All of the ILJ parameters used in this report were improved and tested in comparison with high level *ab initio* calculations as reported in the Ref [34]. Figure 1 shows molecular models that represent the gas molecules. These molecular models along with their point charge distributions are adopted from Ref [41-45], where di- and quadrupole moments of the molecules have been considered.

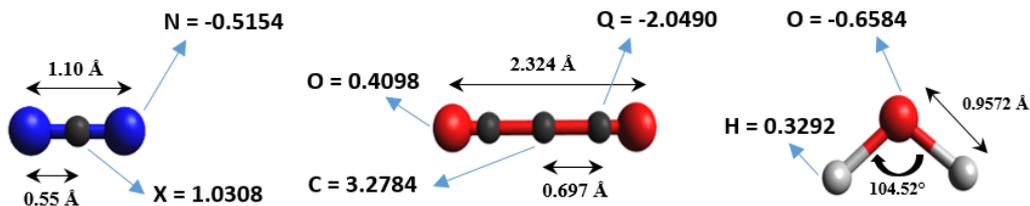

**Fig. 1**. The models used to represent Nitrogen, Carbon dioxide and water molecules along with their point charges. Q is a point charge representing the C–O bond of $CO_2$ while X is a point charge representing the N–N bond of $N_2$.

The MD simulations were performed in a simulation box with dimension 72.210 × 62.523 × 280.0 Å³. The $CO_2/N_2/H_2O$ gaseous mixture with equal fractions of $CO_2$, $N_2$ and $H_2O$ molecules was generated randomly distributing the molecules inside the simulation box. A bilayered-graphtriyne membrane with a dimension of 72.210 × 62.523 Å² was placed in the middle of the box (see Figure 2). The molecular structure of graphtriyne used for the MD simulations is presented in Figure 3. Seven different amounts of gas molecules have been loaded into the box for the simulations, to investigate the influence of pressure in the of gaseous mixtures. The amount of gas molecules inside the box was directly proportional to the initial gas pressure, according to the Peng-Robinson equation of state [46]. In order to mimic post-combustion conditions, four different temperatures (i.e. 333, 353, 373 and 400 K) were considered with initial pressures relatively low pressure, below 5.5 atm.

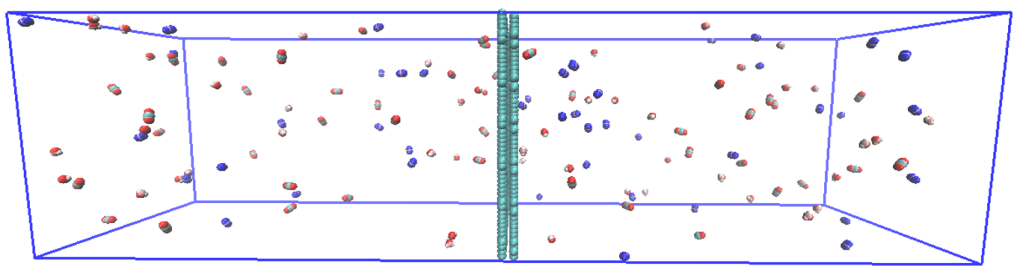

**Fig. 2.** The simulation box was loaded with $CO_2/N_2/H_2O$ gaseous mixtures where the graphytriyne was located at the centre.

The cut-off distances for the ILJ and electrostatic interactions were set to 15 Å, and the Ewald sum method was applied for the calculation of the electrostatic interactions. The graphtriyne layers were considered as a frozen framework and the gas molecules were treated as rigid bodies. Each simulation was performed for 5.5 ns after a 0.5 ns equilibration period with a fixed time step of 1 fs. The statistical data and trajectory were collected at every 2 ps. The gas molecules could cross the layers multiple times in both directions during the simulation. The number of permeation events was then monitored along calculating the $z$-density and the radial distribution function profiles. All the MD simulations were performed by using DL_POLY package [47] in the canonical (NVT) ensemble employing the Nose-Hoover algorithm to maintain the applied temperatures. Periodic boundary conditions were implemented in x, y and z directions. The graphical representations and the molecular trajectories were processed by using the VMD program [48].

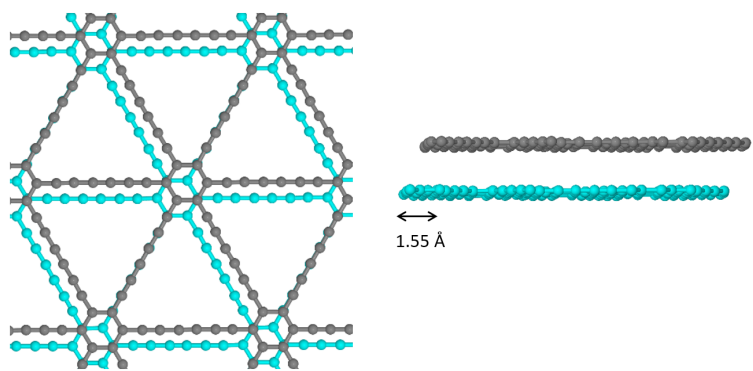

**Fig. 3.** Top and side views of the bilayer graphtriyne structure. According to the structure predicted by the periodic DFT optimization, one of the carbon sheets is shifted of 1.55 Å with respect to the other sheet [33].

## 3. Results and discussions

### *3.1. Gas permeability*

Production runs of each system were started after 0.5 ns equilibration time. The energy and temperature convergence were checked as an indication of the equilibrated system. The number of permeation events occurred during the production runs was counted and plotted as a function of the simulation time. The slope of this plot is an estimation of the gas permeation rate measured in units of molecules ps$^{-1}$. Left panel of Figure 4 shows the permeation events as a function of time for a system simulated at 3.18 atm and 333 K. It can be seen that $CO_2$ has an higher permeation rate than $H_2O$ and $N_2$. The high permeation rate of $CO_2$ can be verified by looking at the radial distribution function profiles that are presented in the right panel of Figure 4. $CO_2$ has the largest and highest peaks meaning that $CO_2$ is more likely to be located near the C atoms of graphtriyne. Consequently, $CO_2$ molecules have the highest probability to permeate the graphtriyne layers. $N_2$ has the lowest permeation due to small number of $N_2$ molecules found near the layers.

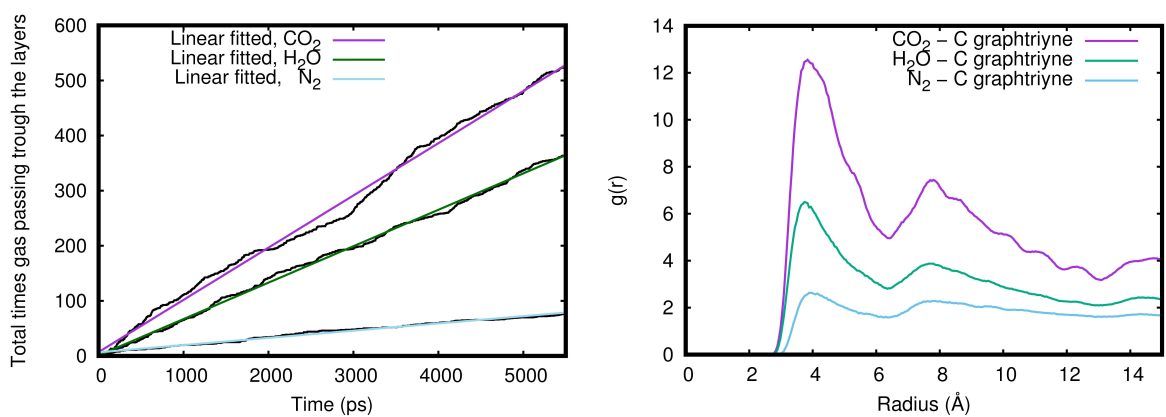

**Fig. 4.** Permeation events (left) and radial distribution functions (right) at 3.18 atm and 333 K.

Furthermore, knowing the permeation rate, the gas permeance can be calculated by dividing it by the corresponding pressure and area of the graphtriyne layers. The gas permeances were reported in gas permeance unit (GPU) (1 GPU = 3.35 × 10$^{-10}$ mol m$^{-2}$ s$^{-1}$ Pa$^{-1}$). In order to widen the applied conditions, we also performed MD simulations at seven different pressures and four different temperatures. The results are reported in Figure 5. Figure 5, left panel, shows that $N_2$ permeances are slightly affected by the pressures. On the other hand, $CO_2$ and $H_2O$ permeances tend to decrease as the pressure increases. From these data, the average of gas permeances were calculated and then plotted as a function of temperature (Figure 5, right panel). It was observed that the average of gas permeance for all gas decreases with the increasing of temperature. This phenomenon is indeed related to their kinetic energy as the temperature varies. The higher is the kinetic energy, the higher is the velocity of molecules to compete with attraction forces of the layers. A high temperature decreases gas permeance by minimizing attraction effects to steer gas molecules toward the graphtriyne layers. The gas permeation reported in Figure 5 is in a good agreement with the potential energy curves reported by Bartolomei and co-workers [33], where well depths increase according to the sequence $N_2$, $H_2O$, $CO_2$ (highest value). Although $CO_2$ has a deeper potential well than $H_2O$, the average gas permeances for $CO_2$ and $H_2O$ are comparable. This fact is closely related with the stereoselective requirement for $CO_2$ to pass through the layers as already discussed in Ref [34]. Nevertheless, by including also the deviation into our consideration (Figure 5), we can say that $CO_2$ still has the highest gas permeance. For the case of $N_2$, low $N_2$ permeance at all temperatures indicates that attraction force of the graphtriyne layers to $N_2$ is too weak compared to that of $CO_2$ and $H_2O$.

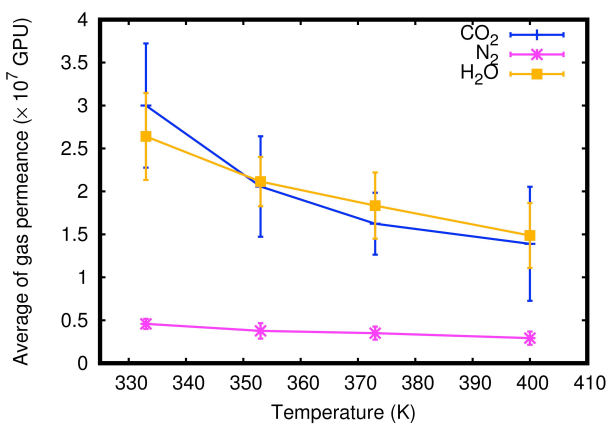

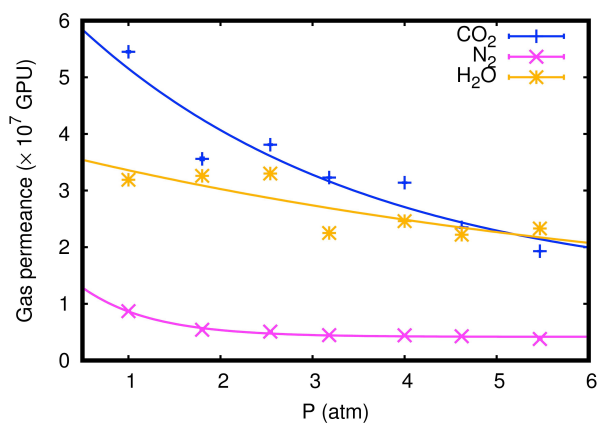

**Fig. 5**. Gas permeance at 333 K (left panel) and average of gas permeance as a function of temperature (right panel).

Although the data presented in Figure 5 follow a similar trend as reported in the Ref [34], the gas permeances in this report are relatively lower (ranging from 0.6 to 3.0 × $10^7$ GPU and 0.3 to 0.6 × $10^7$ GPU for $CO_2$ and $N_2$ respectively). Low values of gas permeances are caused by more competitions existing between the $CO_2$/$N_2$/$H_2O$ gaseous mixtures when interacting with graphtriyne. Moreover, the temperatures applied in this report are also higher. Furthermore, by comparing the average of gas permeances obtained for all type of systems, we calculated permeance selectivities and plotted the values as function of temperature (Figure 6). In general, temperature slightly affects the permeance selectivity of all pairs. Permeance selectivity values of $H_2O$/$N_2$ overlap with those of $CO_2$/$N_2$ by having similar values ranging from 4.8 to 6.5. On the other hand, $H_2O$/$CO_2$ selectivity is around 1 which implies that the bilayer graphtriyne membrane is not selective for $CO_2$−$H_2O$ separation. Although the permeance selectivities are lower than $CO_2$/$N_2$ selectivity reported by Liu *et al.* [15] (about 100, with $CO_2$ permeance = 2.8 × $10^5$ GPU) for nanoporous graphene at 300 K and by Schrier [12] (about 60, with $CO_2$ permeance = 3 × $10^5$ GPU) for Porous Graphene-E-Stilbene-1 (PG-ES1) at 325 K, the $CO_2$ permeances for bilayer graphtriyne (ranging from 0.6 to 3.0 × $10^7$ GPU) are two order of magnitudes higher. Nevertheless, the $CO_2$/$N_2$ permeance selectivities are comparable to those reported by Wu and co-workers [17] for fluorine modified nano-porous graphene at 300 K (ranging from 4 to 11). It is already known that trade-off issue between selectivity and permeability is a drawback of membrane based-materials. However, its simplicity and efficiency make membrane-based technology still widely used for gas separation [18].

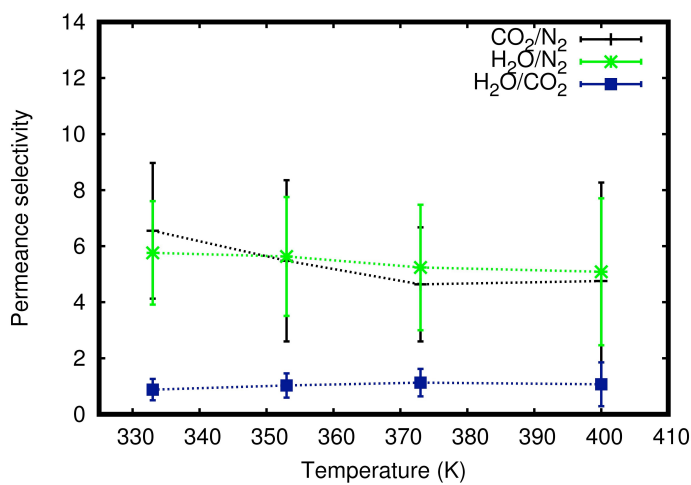

**Fig. 6.** Permeance selectivity as a function of temperature.

### 3.2. *Gas adsorption*

In addition to the radial distribution functions reported in Figure 4, the stronger attraction for $CO_2$ than $H_2O$ and $N_2$ is also reflected in the z-density profiles presented in Figure 7. The z-density is a plot of the mean number density of gas along the z-axis, where the z-axis is perpendicular with the graphtriyne layers. The highest peaks of $CO_2$ indicate that graphtriyne layers exhibit strong adsorption of $CO_2$, especially inside the interlayer region. The z-density profiles also present sharp peaks outside the layers at a distance of around 3.4 Å from the surface for all gas molecules. Figure 7 also shows that the interlayer region is selective for $CO_2$ uptake and separation from the $CO_2/H_2O/N_2$ gaseous mixture. For instance, at 333 K and 4 atm, we obtained interlayer adsorption selectivity of $CO_2/N_2$ and $CO_2/H_2O$ about 20.23 and 1.85 respectively. High and broad peaks of $CO_2$ presented in Figure 7 correspond to the deep and wide of $CO_2$ potential well, meaning a strong long ranged attractions. Therefore, compared with the single layer, the bilayer graphriyne membrane adsorbs more molecules facilitated by their interlayer pores [32]. As already discussed before, the permeation events are closely related to the adsorption of gas over the surfaces of membrane. The more gas is adsorbed over the surfaces, the more probable is the gas to cross the layers. However, the amount of adsorbed molecules in the interlayer region also can diminish the gas permeance by saturating the pores and blocking other molecules to cross the membrane. This phenomenon happens for the case of $CO_2$, where the attraction forces are very strong in the interlayer region. Beside the stereodynamic requirement of $CO_2$ to cross the layers, permeation events of $CO_2$ were also inhibited by other $CO_2$ molecules that occupied the intermolecular pores. Consequently, the average of $CO_2$ permeances are relatively low and comparable with those of $H_2O$ (Figure 5, right panel) even though a lot of $CO_2$ molecules were adsorbed at the surfaces of graphtriyne.

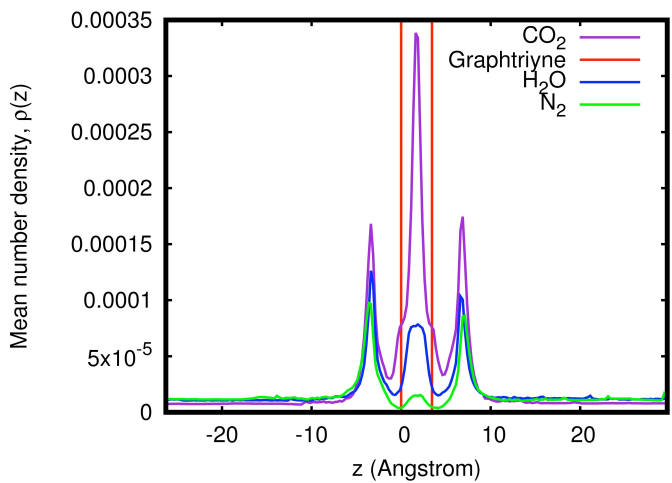

**Fig. 7**. z-density profiles simulated at 1.8 atm and 353 K.

We also calculated the total gas uptakes by integrating the *z*-density profiles at the adsorption region. The adsorption region is an area located in the range of 6.9 Å from the surface of graphtriyne including the area between the layers presented in the z-density profiles. The gas uptakes at 333 K for all applied pressures were presented in the left panel of Figure 8 as adsorption isotherms. It can be seen that the gas uptake is linearly proportional to the initial pressure; the higher is the initial pressure, the higher is the gas uptake value. Furthermore, by calculating the slope of adsorption isotherms, we estimated the adsorption coefficient of each systems. The adsorption coefficients for all molecules are reported in the right panel of Figure 8 as a function of temperature. Figure 8 shows that $CO_2$ has the highest gas uptakes and adsorption coefficients among gas molecules. The adsorption coefficients decrease as the increasing of temperature for all gas molecules. It can be expected that gas physisorption is relatively ineffective at high temperatures. Weak attraction of $N_2$ is manifested by the fact that $N_2$ has the lowest adsorption coefficients (about 0.02 to 0.04 mmol g$^{-1}$ atm$^{-1}$). Low uptake of $N_2$ caused low permeance values of $N_2$ as presented in Figure 5.

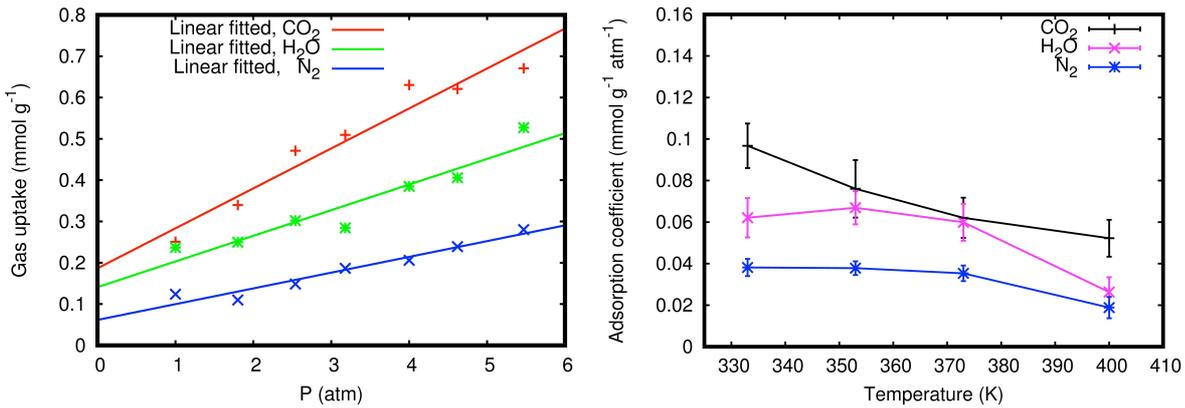

**Fig. 8**. Adsorption isotherm profiles at 333 K (left panel) and adsorption coefficient as a function of temperature (right panel).

Adsorption selectivity of graphtriyne layers were reported in term of total adsorption selectivity. The total adsorption refers to the sum of the interlayer adsorption and the surface adsorption. The total adsorption selectivity $S_{ads}^{A/B}$ were calculated as the following,

$$S_{ads}^{A/B} = \frac{n_{A\,(ads)}}{n_{A\,(free)}} \times \frac{n_{B\,(free)}}{n_{B\,(ads)}}$$

$n_{A\,(ads)}$ and $n_{B\,(ads)}$ are the numbers of adsorbed molecules A and B, while $n_{A\,(free)}$ and $n_{B\,(free)}$ are the numbers of free molecules A and B, respectively. The total adsorption selectivity for all systems are provided in Table 1. In general, total adsorption selectivity of all gas pairs decreases as the increasing of pressure and temperature. Table 1 shows that bilayer graphtriyne exhibits relatively high $CO_2/N_2$ total adsorption selectivity especially at low pressure and temperature. The $CO_2/N_2$ selectivities reported in Table 1 are higher than those reported in our previous report for the single layer system [32]. We obtained $CO_2/N_2$ total adsorption selectivities at 373 K for bilayer system ranging from 2.3 to 5.2, while in the previous work for single layer system are in the range of 1.3 to 2.4 at the same temperature. However, the bilayer graphtriyne is not selective enough for $CO_2/H_2O$ adsorption with total selectivity values ranging from 1 to 2 for all applied temperatures. It has been reported that competitive adsorption between water vapour and $CO_2$ on the same adsorption sites of various materials is observed in post-combustion $CO_2$ capture and separation [5,30]. In most cases, $H_2O$ interferences are minimized by condensing the water molecules or by modifying the adsorbent materials to have more hydrophobic surfaces. Nevertheless, doubling the graphtriyne layer can enhance the amount of adsorbed molecules and the selectivity by providing new adsorption sites in the interlayer pores.

Table 1. Total adsorption selectivity of $CO_2/N_2/H_2O$ gaseous mixture

| Temperature (K) | Gas | Total adsorption selectivity | | | | | | |
|---|---|---|---|---|---|---|---|---|
| | | 1.00 atm | 1.80 atm | 2.54 atm | 3.18 atm | 4.00 atm | 4.62 atm | 5.47 atm |
| 333 | $CO_2/N_2$ | 3.43 | 4.58 | 4.67 | 3.67 | 4.14 | 3.26 | 2.90 |
| | $H_2O/N_2$ | 3.00 | 2.84 | 2.40 | 1.65 | 2.10 | 1.86 | 2.12 |
| | $CO_2/H_2O$ | 1.14 | 1.62 | 1.95 | 2.22 | 1.97 | 1.75 | 1.37 |
| 353 | $CO_2/N_2$ | 3.19 | 4.24 | 3.64 | 4.19 | 2.79 | 2.78 | 2.60 |
| | $H_2O/N_2$ | 2.45 | 1.69 | 2.08 | 2.06 | 1.88 | 1.93 | 1.99 |
| | $CO_2/H_2O$ | 1.30 | 2.51 | 1.75 | 2.03 | 1.48 | 1.44 | 1.31 |
| 373 | $CO_2/N_2$ | 5.16 | 3.04 | 3.40 | 2.67 | 2.62 | 2.62 | 2.33 |
| | $H_2O/N_2$ | 0.81 | 3.12 | 2.05 | 1.42 | 2.04 | 1.59 | 1.84 |
| | $CO_2/H_2O$ | 6.35 | 0.97 | 1.65 | 1.88 | 1.28 | 1.65 | 1.26 |
| 400 | $CO_2/N_2$ | 1.08 | 3.96 | 3.12 | 2.26 | 2.68 | 2.54 | 2.36 |
| | $H_2O/N_2$ | 1.87 | 2.00 | 1.65 | 1.57 | 1.75 | 1.54 | 1.57 |
| | $CO_2/H_2O$ | 0.58 | 1.98 | 1.89 | 1.44 | 1.53 | 1.66 | 1.50 |

## 4. Conclusions

Bilayer graphtriyne has been investigated for the purpose of $CO_2$ capture and separation materials. The gas uptake capacity and permeability of the layers were simulated computationally employing $CO_2/H_2O/N_2$ gaseous mixture systems in wide range of conditions. Extensive MD simulations were performed based on an improved formulation of force fields. Compared to generic force fields, our formulation was tested on accurate *ab initio* calculations on specific systems for mixture separation and gas capture in post-combustion process. Therefore, a quantitative description of the interactions and realistic results were obtained for the dynamics under considered systems. We observed that the bilayer graphtriyne membrane has high $CO_2$ and $H_2O$ permeances compared to $N_2$ with relatively good selectivity (from 4.8 to 6.5). Adsorbed $CO_2$ lowers the $CO_2$ permeance by occupying the interlayer pores and preventing permeation events. As a result, $CO_2$ has permeance values similar to $H_2O$, despite the $CO_2$ adsorption is favoured than other molecules. High $CO_2$ uptake capacity and selectivity were founded in the interlayer region. The interlayer adsorption selectivity of $CO_2/N_2$ and $CO_2/H_2O$ are about 20.23 and 1.85, respectively at 333 K and 4 atm. The layers are not selective enough for $CO_2/H_2O$ capture and separation as also reported for other materials. Nevertheless, in post-combustion application, competitive $CO_2/H_2O$ adsorption can be reduced by condensing $H_2O$ prior interacting the flue gas with the graphtriyne layers. In addition, we also investigated the effect of temperature to the gas uptake capacity, permeability and selectivity. As the temperature increases; the gas uptake, adsorption coefficient and gas permeance decrease for all gas molecules. Among carbon-based materials, bilayer graphtriyne can be considered as an alternative membrane with its potential properties for mixture separation that offers fast and efficient separation of different molecular species.

## Acknowledgements

YBA thanks to the LCPQ - Université de Toulouse III for allocated computing time. N. F.-L and A. L. thanks MIUR and the University of Perugia for the financial support of the AMIS project through the "Dipartimenti di Eccellenza" programme. N. F.-L. also acknowledges the Fondo Ricerca di Base 2017 (RICBASE2017BALUCANI) del Dipartimento di Chimica, Biologia e Biotecnologie della Università di Perugia for financial support. A. L. acknowledges financial support from MIUR PRIN 2015 (contract 2015F59J3R 002) and the OU Super-computing Center for Education & Research (OSCER) at the University of Oklahoma, for allocated computing time.